\pdfoutput=1
\documentclass[conference,a4paper,10pt]{IEEEtran}
\usepackage[keeplastbox]{flushend}
\usepackage[latin1]{inputenc}
\usepackage{times,amsmath}
\usepackage{amsfonts}
\usepackage{pstool}
\usepackage{subfigure}
\usepackage{multirow}
\usepackage{multicol}
\usepackage{enumerate}
\usepackage{graphicx}
\usepackage{mathtools, cuted}
\usepackage{MnSymbol}
\usepackage{stfloats}
\usepackage[table]{xcolor}
\usepackage[square, comma, sort&compress, numbers]{natbib}
\usepackage{nohyperref}
\usepackage{algorithm,algorithmic}
\usepackage{bigints}
\usepackage{amsmath}

\newcounter{tempEquationCounter}
\newcounter{thisEquationNumber}

\makeatletter
\newcommand{\vast}{\bBigg@{4}}
\newcommand{\Vast}{\bBigg@{5}}
\newcommand\numeq[1]%
  {\stackrel{\scriptscriptstyle(\mkern-1.5mu#1\mkern-1.5mu)}{=}}
\makeatother

\graphicspath{ {Figures/} }

\begin{document}


\title{On the Effective Capacity of IRS-assisted Wireless Communication}

\author{
\IEEEauthorblockN{Waqas\ Aman$^{\ast \dagger} $, M.\ Mahboob\ Ur\ Rahman$^\ast$, Shuja Ansari$^\dagger$, Ali Arshad Nasir$^\ddagger$, Khalid Qaraqe$^\bot$,\\ M.\ Ali\ Imran$^\dagger$,  Qammer\ H.\ Abbasi$^\dagger$\\
$^\ast$Electrical engineering department, Information Technology University, Lahore 54000, Pakistan \\
$^ { \dagger}$Department of Electronics and Nano Engineering, University of Glasgow, Glasgow, G12 8QQ, UK \\
$^ {\ddagger}$Department of Electrical Engineering, King Fahd University of Petroleum and Minerals, Dhahran 31261, Saudi Arabia\\
$^ {\bot}$Electrical and computer engineering, Texas A \& M University at Qatar\\
$^\ast$\{waqas.aman, mahboob.rahman\}@itu.edu.pk, $^\dagger$\{Shuja.Ansari, Muhammad.Imran, Qammer.Abbasi\}@glasgow.ac.uk \\ $^\ddagger$anasir@kfupm.edu.sa $^\bot$khalid.qaraqe@qatar.tamu.edu.
}
}

\maketitle


\maketitle

\begin{abstract}

We consider futuristic, intelligent reflecting surfaces (IRS)-aided communication between a base station (BS) and a user equipment (UE) for two distinct scenarios: a single-input, single-output (SISO) system whereby the BS has a single antenna, and a multi-input, single-output (MISO) system whereby the BS has multiple antennas. For the considered IRS-assisted downlink, we compute the effective capacity (EC), which is a quantitative measure of the statistical quality-of-service (QoS) offered by a communication system experiencing random fading. For our analysis, we consider the two widely-known assumptions on channel state information (CSI)---i.e., perfect CSI and no CSI, at the BS. Thereafter, we first derive the distribution of the signal-to-noise ratio (SNR) for both SISO and MISO scenarios, and subsequently derive closed-form expressions for the EC under perfect CSI and no CSI cases, for both SISO and MISO scenarios. Furthermore, for the SISO and MISO systems with no CSI, it turns out that the EC could be maximized further by searching for an optimal transmission rate $r^*$, which is computed by exploiting the iterative gradient-descent method. We provide extensive simulation results which investigate the impact of the various system parameters, e.g., QoS exponent, power budget, number of transmit antennas at the BS, number of reflective elements at the IRS etc., on the EC of the system. 
\\
\end{abstract}


\section{Introduction}
\label{sec:intro}
An intelligent reflecting surface (IRS) (also known as re-configurable intelligent surface or smart radio-environment) consists of a group of passive, re-configurable meta-surface elements. Guided by a controller, the IRS elements together could steer the incident radio frequency waves in any arbitrary direction by changing their individual reflective angles \cite{ozdogan:WCL:2019}. Due to the easy reconfiguration for transmit beamforming, ease of deployment, low hardware complexity and low power consumption, IRS is anticipated to be one of the core components of the beyond 5G/6G cellular networks \cite{Gong:Arxiv:2019}, \cite{zhao:arXiv:2019}, \cite{huang:arXiv:2020}. Moreover, IRS is foreseen to enable a number of novel application scenarios, e.g., high-fidelity communication at the cell-edge and in an area which has become a dead zone due to blockage, massive device-to-device communication, design of novel schemes for physical layer security, wireless power transfer etc. \cite{huang:arXiv:2020}, \cite{wu:CM:2020}. 
 
IRS-assisted communication networks have attracted lot of attention during the last few years; and thus, an exhaustive discussion of all the prior work on IRS-assisted systems is beyond the scope of this work (to this end, the interested reader is referred to the survey article \cite{Gong:Arxiv:2019}, and the references therein). Therefore, only selected relevant works are summarized here. \cite{Huang:TWC:2019} proposes IRS-assisted communication to achieve higher energy efficiency, while \cite{Schober:arXive:2019} exploits the architecture of the considered IRS-assisted system to design novel schemes for physical layer security. \cite{Wu:GLOBECOM:2018} solves the problem of joint active beamforming (by the transmitter) and passive beamforming (by the IRS) for maximization of the receiver power at the receiver. \cite{zhao2:ARXiv:2020} studies a realistic scenario whereby the transmitter has the perfect knowledge of the channel and does active beamforming, while the IRS utilizes the statistical knowledge of the channel (e.g., pathloss information) to do passive beamforming. \cite{Zhou:TSP:2020} considers an IRS-aided multi-user system and studies the design of robust reflection beamforming in the face of imperfect knowledge of all the cascaded channels under consideration, for two different models of the channel estimation error. Note that a cascade channel represents the cumulative impact of the environment as the signal traverses from the transmitter to an IRS element, and then from that IRS element to the user. In \cite{zhao:ARXiv:2020}, authors consider an IRS-aided multi-user system and study the impact of the imperfect channel knowledge on the joint design of the transmit beamforming at the transmitter and the reflection coefficients (with both phase control and amplitude control) at the IRS in order to maximize the weighted sum rate of the system. Last but not the least, \cite{Huang:JSAC:2020} considers a multi-user system and exploits deep reinforcement learning to jointly design the beamforming matrix at the transmitter and the phase shift matrix at the IRS. 
 
Effective capacity (EC), on the other hand, is the maximum-sustainable constant arrival rate at a transmitter's queue in the face of randomly time-varying service process \cite{Wu:TWC:2003}. In other words, the EC is a quantitative measure of the throughput of a wireless fading channel under statistical quality-of-service (QoS) constraints. So far, the researchers have computed the EC for a wide variety of communication systems under different assumptions and various channel conditions, e.g., cognitive radio channels \cite{Gursoy:TWC:2010}, \cite{Anwar:TVT:2016}, systems with various degrees of channel state information (CSI) at the transmitter \cite{gross2012scheduling}, two-hop systems \cite{Gursoy:TIT:2013}, \cite{Lateef:TC:2009}, correlated fading channels \cite{Soret:TWC:2010}, device-to-device communication \cite{WShah:WCL:2019}, and underwater acoustic communication \cite{waqas:ICC:2020}. 

{\bf Contributions.} 
This work investigates the statistical QoS offered by an IRS-aided communication between a base station (BS) and a user equipment (UE) for two distinct scenarios: a single-input, single-output (SISO) system whereby the BS has a single antenna, and a multi-input, single-output (MISO) system whereby the BS has multiple antennas.
The contribution of this work is two-fold: 
\begin{itemize}
\item EC analysis of the IRS-assisted SISO system: 
We begin with exploiting the central-limit theorem to derive the distribution of the signal-to-noise ratio (SNR) at the UE. Then, for the case of perfectly-known CSI at the BS, we derive the EC by computing the log moment generating function (LMGF) of the service process via integration-by-parts. For the case of no CSI at the BS, we model the underlying dynamical system as Markov chain with state-transition probability matrix $\mathbf{P}$, and derive the EC by computing the spectral radius of a special block-companion matrix $\mathbf{P\Theta}$ where $\mathbf{\Theta}$ is a diagonal reward matrix. Additionally, for the no CSI case, the EC is further maximized by computing the optimal transmission rate via iterative gradient-descent method. 
\item EC analysis of the IRS-assisted MISO system: 
We first derive the SNR distribution. Then, for the case of perfectly-known CSI at the BS, we compute the EC by computing the first-order and the second-order moments of the service process at hand. For the case of no CSI at the BS, we again do Markov chain based analysis to compute the EC as the spectral radius of a special block-companion matrix $\mathbf{P\Theta}$. Again, for the no CSI case, we provide a closed-form expression which is an approximation of the optimal transmission rate. 
\end{itemize}

{\bf Outline}: The rest of this paper is organized as follows. Section-II presents the system model, and introduces the notion of of the EC. Section-III computes the EC for the SISO scenario under the assumptions of perfect CSI and no CSI at the BS. Section-IV computes the EC for the MISO scenario, again under the assumptions of perfect CSI and no CSI at the BS. Section-V presents extensive simulation results. Section-VI concludes the paper.

{\bf Notations}: Bold-face small letters (e.g., $\mathbf{a}$) represent vectors, while bold-face capital letters (e.g., $\mathbf{A}$) represent matrices. $||a||$ represents the 2-norm of vector $\mathbf{a}$, $(\mathbf{a})^H$ represents the hermitian of vector $\mathbf{a}$. $A \sim \mathcal{CN}(0,1)$ implies that $A$ is a random variable with complex normal distribution. $E(.)$ represents the statistical expectation operator.

\section{System Model \& Background}

\subsection{System Model}

This work considers a two-hop downlink whereby a BS communicates with a UE through a panel of $N$ passive, reflective IRS elements (see Fig. 1). 

We make the following assumptions. A1) The direct path between the BS and the UE doesn't exist due to blockage; and hence, there are two hops (from the BS to the IRS, and from the IRS to the UE). A2) The channel fading/service process elements across the slots on both hops are block fading channels, and are independent and identically distributed (i.i.d.). A3) Perfect CSI is always available for both hops at the IRS controller. A4) The system under consideration is a time-slotted system with $T$ seconds long time-slots.

Having said that, we first consider the SISO scenario (whereby the BS has a single antenna), followed by the MISO scenario (whereby the BS has $N_t$ antennas). 

\begin{figure}[ht]
\begin{center}
	\includegraphics[width=3.0in,height=1.8in]{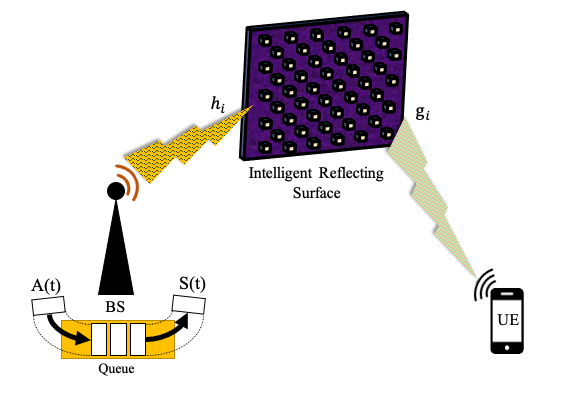}
\caption{The system model: we examine the QoS performance of a downlink channel whereby a panel of $N$ passive, reflective IRS elements assists the communication between the BS and the UE. $A(t)$, $S(t)$ represent the arrival process and the service process at the BS queue, respectively.}
\label{fig:ann}
\end{center}
\end{figure}

{\bf SISO scenario:}
Let $p_t$ be the total transmit power budget of the BS. Let $x$ be the symbol that is transmitted by the BS, then the received signal $y$ at the UE is given as:

\begin{align}
\label{eq:y}
y=\sqrt{p_t}.\sqrt{\zeta} \sum_{i=1}^Nh_ig_ie^{j\phi_i}x+w
\end{align}
where $h_i \sim \mathcal{CN}(0,1)$ represents the channel gain from the BS to the $i$-th element of the IRS, and $g_i \sim \mathcal{CN}(0,1)$ represents the channel gain from the $i$-th element of the IRS to the UE (see Fig. 1). Moreover, $\phi_i \in [0,2\pi)$ is the phase control employed by the controller of the IRS. Furthermore, $\zeta=\frac{G_t G_r}{(4\pi)^2} (\frac{x_{IRS}y_{IRS}}{d_1 d_2})^2 \cos^2(\varphi)$ is the pathloss between the BS and the UE through the IRS \cite{ozdogan:WCL:2019}. In the preceding pathloss expression, $d_1$ represents the distance between the BS and the IRS; $d_2$ represents the distance between the IRS and the UE; $\varphi \in [0,\pi/2]$ represents the angle of incidence at the IRS; $x_{IRS}$,$y_{IRS}$ represent the length and the width of the IRS respectively; $G_t$, $G_r$ represent the antenna gain at the BS and at the UE, respectively. Finally, $w \sim \mathcal{CN}(0,\sigma^2)$ is the zero-mean, additive white Gaussian noise (AWGN) at the UE with variance $\sigma^2$.  

 {\it Proposition 2.1:} The SNR at the UE (during slot $n$) is:  $\gamma(n) = \beta \chi_1^2(\lambda)$, where $\chi_1^2(\lambda)$ is a non-central chi-square random variable with one degree-of-freedom and non-centrality parameter $\lambda=\frac{N\pi^2}{(16-\pi^2)} >0$. Moreover, $\beta=N\frac{p_t\zeta(16-\pi^2)}{4\sigma^2}$.
 
 {\it Proof:} Given in Appendix A. 
 
Next, with the SNR $\gamma(n)$ in hand, one can compute the Shannon capacity of the SISO downlink channel under consideration (during slot $n$) as follows:
\begin{align}
c(n)= B \ \log_2(1+\gamma(n)),
\end{align}
where $B$ is the bandwidth of the system.

{\bf MISO scenario:}
The BS transmits the same symbol $x$ from all the $N_t$ transmit antennas (in order to achieve antenna diversity). Then, the received signal $y$ at the UE is given as:
\begin{align}
y=\sqrt{p_t}\sqrt{\zeta}\mathbf{g}^H \mathbf{\Pi}\mathbf{Hf}x+w ,
\end{align}
where $\mathbf{g}^H=[g_1 \ .\ .\ g_N]$ , $\mathbf{\Pi}=diag[e^{j\phi_1}  \ . \ .\  e^{j\phi_N}]$ is the so-called reflection coefficients matrix of dimension $N\times N$, $\mathbf{H}$ is the $N \times N_t$ channel matrix between the BS and the IRS, and $\mathbf{f}=[f_1 \ .\ .\ f_{N_t}]^T$ is the transmit precoding vector at the BS. 

{\it Proposition 2.2:} The SNR at the UE (during slot $n$) is: $\gamma(n)=p_t\zeta \frac{\vert \mathbf{g}^H\mathbf{\Pi}\mathbf{Hf} \vert^2}{\sigma^2}$. Specifically, $\gamma (n)$ is an exponential random variable with parameter $\kappa$.  

{\it Proof:} Given in Appendix B.

Next, with the SNR $\gamma(n)$ in hand, one can compute the Shannon capacity of the MISO downlink channel under consideration (during slot $n$) as: $c(n)= B \ \log_2(1+\gamma(n))$.

\subsection{Background: Effective Capacity}

For a stationary and ergodic channel (service process), the effective capacity (EC) provides the maximum-sustainable constant source rate at the queue of a transmitter, and is given as the limit \cite{Wu:TWC:2003}:
\begin{align}
\label{eq:EC}
EC(\alpha)= -\lim_{t\rightarrow \infty} \frac{1}{\alpha t} \ln(E(e^{-\alpha S(t)}))=-\frac{\Lambda(-\alpha)}{\alpha} 
\end{align}
where $\Lambda(\alpha)=\lim_{t\rightarrow \infty} \frac{1}{t} \ln(E(e^{\alpha S(t)}))$ is the Gartner-Ellis limit on the service process given as the LMGF of the cumulative service process $S(t)=\sum_{n=1}^t s(n)$ where $s(n)$ is the service process (number of bits successfully served) during time-slot $n$. Moreover, $\alpha >0$ is the so-called QoS exponent. Specifically, when $\alpha$ tends to zero, it implies delay-tolerant communication. On the other hand, when $\alpha$ tends to infinity, it implies delay-limited communication. Note that the EC in Eq. \ref{eq:EC} has the units of bits/block.

In short, the notion of the EC allows us to determine the maximum fixed source rate under a statistical QoS constraint---$P(D>D_m)\leq \epsilon$, where $D$ is the steady-state delay experienced by the packets at the queue of the BS, $D_m$ is the delay target, and $\epsilon$ is the delay-violation probability.

\section{EC analysis for IRS-assisted SISO Downlink}
\label{sec:sys-model}

We first compute the EC of an IRS-assisted SISO downlink for the case when the BS has perfect knowledge of the CSI, followed by the case when the BS has no knowledge of the CSI.

\subsection{Perfect CSI at the BS}
When the BS has perfect CSI available, it transmits at the Shannon rate $c(n)$ during slot $n$, which implies that $s(n)=c(n)$. Moreover, due to the assumption A2 which states that the block fading is i.i.d. on both hops (and thus, the service process elements $s(n)$ are i.i.d.), Eq. \ref{eq:EC} could be simplified as follows\footnote{The EC is defined as: $EC=\lim_{t\rightarrow \infty} -\frac{1}{\alpha t} \ln(E(e^{-\alpha \sum_{n=1}^ts(n)}))$. This expression could be re-written as: $EC = \lim_{t\rightarrow \infty} -\frac{1}{\alpha t} \ln( \Pi_n E(e^{-\alpha s(n)}))$. Using the identity $\ln(a.b)=\ln(a)+\ln(b)$, we could re-write EC as follows: $EC = \lim_{t\rightarrow \infty} -\frac{1}{\alpha t} \sum_n \ln( E(e^{-\alpha s(n)}))$. Finally, due to the fact that the service process elements $s(n)$ are i.i.d., EC is further simplified as follows: $EC = -\frac{1}{\alpha } \ln(E(e^{-\alpha s(1)}))$, where $s(1)=c(1)$.} \cite{Tang:TWC:2007}:
\begin{align}
\label{EC:pCSIT}
EC =  -\frac{1}{\alpha } \ln(E(e^{-\alpha s(1)}))
\end{align}
where $s(1)=c(1)$. Next, we compute the moment generating function (MGF) of the service process $s(1)$ during slot 1 as follows:

\begin{align}
E(e^{-\alpha s(1)})=\int_0^\infty e^{-\alpha B \log_2(1+\beta x)}f_X(x) dx
\end{align}
where $f_X(x)$ is the probability density function (pdf) of $X \sim \chi_1^2(\lambda)$. Next, we do the following relaxation: $(1+\beta x)\approx (\beta x)$. After some simplification, and utilizing a result (Theorem $3.4.1$) from \cite{Murihead:Wiley:1982}, above equation could be re-written as follows:
\begin{align}
\label{eq:intbp}
E(e^{-\alpha s(1)})=\frac{e^{\frac{-\lambda}{2}}}{2^{\frac{1}{2}}\Gamma(\frac{1}{2})\beta^{\frac{\alpha B }{\ln(2)}}}\int_0^\infty  \prescript{}{0}{\mathbf{F}}_1(;\frac{1}{2};\frac{\lambda x}{4})\frac{e^{-\frac{x}{2}}}{(x)^{\frac{\alpha B }{\ln(2)}+\frac{1}{2}}} dx
\end{align}
where $\Gamma (.)$ is the gamma function and $\prescript{}{x}{\mathbf{F}}_x(.,.,.)$ is the generalized hypergeometric function. Next, using the fact that $\int_{0}^\infty x^{a-1}e^{-bx}\prescript{}{0}{\mathbf{F}}_1(;,c,x) dx = b^{-a}\Gamma(a)\prescript{}{1}{\mathbf{F}}_1(a,c,\frac{1}{b})$, and after some simplifications, Eq. \ref{eq:intbp} can be written as:
\begin{align}
\label{eq:intbp2}
&E(e^{-\alpha s(1)})=\frac{e^{\frac{-\lambda}{2}}}{2^{\frac{1}{2}}\Gamma(\frac{1}{2})\beta^{\frac{\alpha B }{\ln(2)}}(\frac{\lambda}{4})^{\frac{\alpha B }{\ln(2)}+\frac{1}{2}}}\times \nonumber \\ &\left[ (\frac{2 \lambda }{4})^{\frac{\alpha B }{\ln(2)}+\frac{3}{2}} \Gamma (\frac{\alpha B }{\ln(2)}+\frac{3}{2}) \prescript{}{1}{\mathbf{F}}_1(\frac{\alpha B }{\ln(2)}+\frac{3}{2},\frac{1}{2},\frac{2 \lambda }{4})   \right] 
\end{align}
 Finally, putting Eq. \ref{eq:intbp2} into Eq. \ref{EC:pCSIT} leads us to the final expression of the EC. 
  
  \subsection{No CSI at the BS}
\begin{figure}[ht]
\begin{center}
	\includegraphics[width=3.8in]{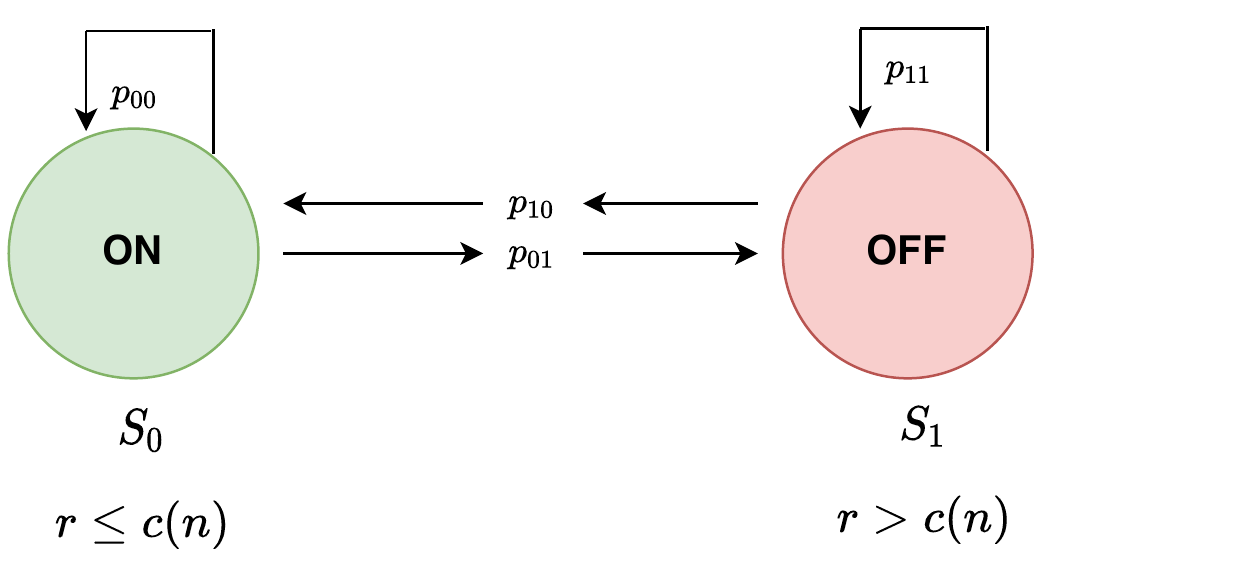} 
\caption{Markov chain representation of the IRS-aided SISO downlink for the case of no CSI at the BS. }
\label{fig:MC}
\end{center}
\end{figure}

  We now compute the EC of the IRS-aided SISO downlink when there is no CSI knowledge available at the BS. In this case, rate adaptation by the BS is not possible; thus, it transmits at a constant rate of $r$ bits/sec during every slot. Then, due to Shannon's channel capacity limit $c(n)$ (see Eq. 2), the SISO downlink under consideration becomes an ON-OFF channel. Thus, it could be modelled as a Markov chain with two states, namely $S_0$ (ON state) and $S_1$ (OFF state). That is, when $r \leq c(n)$, then the channel is ON ($rT$ bits are received by the UE), and is in state $S_0$. Here, $T$ represents the duration of a time-slot. On the other hand, when $r > c(n)$, then the channel is OFF (zero bits are received by the UE), and is in state $S_1$. The motivation behind Markov chain modelling of the SISO downlink comes due to a well-reputed result \cite{Chang:TNCS:2012} which states that one viable way to compute the EC is by computing the spectral radius of a special block-companion matrix $\mathbf{P\Theta}$ where $\mathbf{P}$ is the so-called state-transition probability matrix and $\mathbf{\Theta}$ is a diagonal reward matrix. The Markov chain representation of the IRS-aided SISO downlink (for the scenario of no CSI at the BS) is shown in Fig. \ref{fig:MC}. The entries of the matrix $\mathbf{P}$ for the two-state Markov chain at hand are also labelled in Fig. 2.
  
  The state-transition probability matrix $\mathbf{P}$ of the Markov chain under consideration is formally defined as:
  \begin{align}
\mathbf{P}=  
  \begin{bmatrix}
p_{00} & p_{01} \\
p_{10} & p_{11}
\end{bmatrix}
  \end{align}
where $p_{xy}$ is the state-transition probability from the state $S_x$ during slot $n-1$ to the state $S_y$ during slot $n$. 

Let's compute all the four entries of $\mathbf{P}$ one by one. To this end, the probability $p_{00}$ is defined as:
\begin{align}
p_{00} = P(r<c(n)\mid r<c(n-1))
\end{align}
Next, recall the assumption A2 which states that block fading channels are i.i.d. across the slots; thus, the event $r<c(n-1)$ is statistically independent from the event $r<c(n)$. In other words, the stochastic process $\{c(n)\}$ (or, equivalently, the stochastic process $\{\gamma(n)\}$) is a memoryless process. Therefore, if the SISO downlink at hand has evolved to, say, the state $S_0$ during slot $n$, then the system is completely oblivious/indifferent of the state ($S_0$ or $S_1$) it was into during the previous slot $n-1$. This allows us to lead to the conclusion that the elements of the first column of $\mathbf{P}$ are identical: $p_{00}=p_{10}=p_{0}$. Similarly, the elements of the second column of $\mathbf{P}$ are identical: $p_{01}=p_{11}=p_{1}$. Thus, the matrix $\mathbf{P}$ becomes a unit-rank matrix:
\begin{align}
\mathbf{P}=  
  \begin{bmatrix}
p_{0} & p_{1} \\
p_{0} & p_{1}
\end{bmatrix}
  \end{align}

Next, we compute $p_0$ as follows:
\begin{align}
p_0=P(r<c(n))=P(\gamma(n)>\frac{2^{\frac{r}{B}}-1}{\beta})=Q_{\frac{1}{2}}(\sqrt{\lambda}, {\Xi})
\end{align}
where $Q_x(.,.)$ is the Marcum Q-function of order $x$, and $\Xi = \sqrt{\frac{2^{\frac{r}{B}}-1}{\beta}}$. Similarly,
\begin{align}
p_1=1-Q_{\frac{1}{2}}(\sqrt{\lambda}, {\Xi})
\end{align}
Next, we utilize the findings of \cite{Chang:TNCS:2012} to compute the EC as follows:
\begin{align}
EC = -\frac{1}{\alpha}\ln(sp(\mathbf{P}\mathbf{\Theta}(-\alpha)))
\end{align} 
where $sp(\mathbf{A})$ computes the spectral radius of the matrix $\mathbf{A}$, and $\mathbf{\Theta}$ is a diagonal matrix which contains the MGFs of the service processes for the two states ($S_0$ and $S_1$) on its main diagonal. In state $S_0$, $rT$ bits are received at the UE during a given time-slot; therefore, the MGF of the service process is: $e^{\alpha rT}$. On the other hand, when the system is in state $S_1$, zero bits are received at the UE; therefore, the MGF of the service process is: $e^0=1$. Thus, the final expression of the EC is as follows: 
\begin{align}
\label{eq:EC_SISO_NC}
EC = -\frac{1}{\alpha}\ln(p_0e^{-\alpha rT}+p_1)
\end{align}

{\bf Computation of the optimal transmission rate.}
For the scenario of no CSI at the BS, the EC could be maximized further by searching for the optimal transmission rate $r^*$ at the BS. This is because we note that Eq. \ref{eq:EC_SISO_NC} is a concave-like function of $r$. Thus, one could further enhance the EC in Eq. \ref{eq:EC_SISO_NC} by searching for an optimal transmission rate. This is achieved by taking the derivative of Eq. \ref{eq:EC_SISO_NC}  w.r.t. $r$ and equating it to zero. In other words, $r^\ast = \arg \max_r EC(r)$. Moreover, Eq. \ref{eq:EC_SISO_NC} implies that the maximizing the EC is equivalent to minimizing the argument of the $\ln(.)$ on its right-hand side. Therefore,
\begin{align}
   r^\ast = \arg \max_r EC(r) = \arg \min_r( p_0e^{-\alpha rT}+p_1)  
\end{align}
Taking partial derivative of $\rho (r)= (p_0e^{-\alpha rT}+p_1)$ w.r.t. $r$, we get:
\begin{align}
\label{eq:rSISO}
 \frac{\partial \rho}{\partial r}= Q_{\frac{1}{2}}^\dagger (\sqrt{\lambda}, \Xi)\frac{e^{-\alpha rT} \frac{\ln (2)}{B} 2^{r/B}}{2\sqrt{\beta(2^{\frac{r}{B}}-1)}}+&e^{-\alpha rT}({-\alpha T})Q_{\frac{1}{2}} (\sqrt{\lambda}, \Xi) - \nonumber\\&Q_{\frac{1}{2}}^\dagger (\sqrt{\lambda}, \Xi)\frac{\frac{\ln (2)}{B} 2^{r/B}}{2\sqrt{\beta(2^{\frac{r}{B}}-1)}}
\end{align}
where $Q_{\frac{1}{2}}^\dagger (\sqrt{\lambda}, \Xi)=\frac{\partial Q_{\frac{1}{2}} (\sqrt{\lambda}, \Xi) }{\partial \Xi} = -\frac{\Xi^{\frac{1}{2}}}{\lambda^{\frac{-1}{4}}}e^{-\frac{(\lambda+\Xi^2)}{2}}I_{\frac{-1}{2}}(\sqrt{\lambda}\Xi)$ \cite{Yu:2012}, where $I_x(.)$ is the modified Bessel function of order $x$. Now, equating Eq. \ref{eq:rSISO}
to zero reveals that solving for $r^\ast$ is quite involved due to the presence of infinite number of terms in both the modified Bessel function as well as the Marcum Q-function. Therefore, we propose an alternative approach where we compute $r^\ast$ using an iterative, gradient-descent method. Specifically, the gradient-descent method begins with an initial guess $r(0)$ of the transmission rate, and utilizes the following control law to refine its guess of $r$ during iteration $m$:
\begin{align}
r(m)=r(m-1)- \delta \frac{\partial \rho}{\partial r}\Bigg|_{r=r(m-1)}
\end{align}
where $\delta$ is the step size (basically, an algorithm parameter), and $\frac{\partial \rho}{\partial r}$ is the gradient of the EC function (see Eq. \ref{eq:rSISO}) that gets evaluated at $r=r(m-1)$. The algorithm terminates when the following condition is met: $r(m)-r(m-1)\leq \epsilon_c$, where $\epsilon_c$ is a small constant. If the algorithm terminates, say, during iteration $\mathcal{M}$, it outputs: $r^\ast = r(\mathcal{M})$.

\section{EC analysis for IRS-assisted MISO Downlink}
\label{sec:MISO}

We first compute the EC of an IRS-assisted MISO downlink for the case when the BS has perfect knowledge of the CSI, followed by the case when the BS has no knowledge of the CSI.

\subsection{Perfect CSI at the BS}
Recall that we assume the channel fading/service process elements across the slots on both hops to be i.i.d. Moreover, we assume that the time-window $t$ under consideration is reasonably large so that one could invoke the Central limit theorem to approximate the distribution of the cumulative service process $S(t)=\sum_{n=1}^t s(n)$ as Gaussian \cite{Soret:TWC:2010}. Thus, the EC expression in Eq. \ref{eq:EC} becomes the LMGF of a Gaussian random variable, and is given as:
\begin{align}
\label{eq:ECMISOPCSI}
    EC =\mu_{s(1)}- \frac{\alpha}{2}\sigma_{s(1)}^2
\end{align}
where $\mu_{s(1)}$ is the mean of service process $s(1)$, while $\sigma_{s(1)}^2$ is the variance of the service process $s(1)$. In other words, to compute the EC, we need to find the first and second moments of the service process. The first moment (mean) is given as:
\begin{align}
    \mu_{s(1)}= B \int_{0}^\infty \log_2(1+x)f_{X}(x)dx
\end{align}
where $f_{X}(x)=\kappa e^{-\kappa x}$ is the pdf of $X \sim \text{exp}(\kappa)$. Recall that $X$ is the SNR at the UE as defined in Proposition 2.2. In terms of natural-log, the above equation could be expressed as:
\begin{align}
\mu_{s(1)}=\frac{B\kappa}{\ln(2)} \int_{0}^\infty \ln(1+x)e^{{-\kappa x}}dx
\end{align}
Applying integration-by-parts, we have:
\begin{align}
\label{eq:firstmoment}
\mu_{s(1)} &=\frac{B\kappa}{\ln(2)}\left[\ln(1+x)e^{-\kappa x}\frac{-1}{\kappa}+\frac{1}{\kappa}\int \frac{e^{-\kappa x}}{1+x}dx \right]_0^\infty \nonumber \\
&=\frac{B}{\ln(2)} e^{\kappa} E_1(\kappa)
\end{align}
where $E_1(.)$ is the exponential integral function. 

Next, the second moment $\eta_{s(1)}$ of the service process $s(1)$ could be computed as follows:
\begin{align}
\eta_{s(1)} = \frac{B^2\kappa}{\ln^2(2)} \int_{0}^\infty \ln^2(1+x)e^{{-\kappa x}}dx 
\end{align}
We follow the steps of \cite{gross2012scheduling} to compute the second moment and the final result is given as:
\begin{align}
\label{eq:secondmoment}
\eta_{s(1)} = & \frac{B^2e^{{\kappa}}}{\ln^2(2)}\left[\frac{\pi^2}{6}+C^2+2C\ln({\kappa})+\ln^2({\kappa})\right]- \nonumber \\   &\frac{2 \kappa B^2 e^{{\kappa}}}{\ln^2(2)}  \prescript{}{3}{\mathbf{F}}_3([1,1,1],[2,2,2], -\kappa) 
\end{align}
where  $C$ is the Euler constant. 

Thus, $\sigma_{s(1)}^2$ could be computed as follows: $\sigma_{s(1)}^2 = \eta_{s(1)} - \mu_{s(1)}^2$ where $\eta_{s(1)}$, $\mu_{s(1)}$ are given in Eqs. \ref{eq:secondmoment} and \ref{eq:firstmoment}, respectively. Finally, plugging the values of $\sigma_{s(1)}^2$ and $\mu_{s(1)}$ into Eq. \ref{eq:ECMISOPCSI} lead us to final expression of the EC. 

\subsection{No CSI at the BS}

We now compute the EC of the IRS-aided MISO system when there is no CSI knowledge at the BS. As we will see, the conceptual framework is the same as the IRS-aided SISO system with no CSI at the BS. That is, the BS transmits at a fixed rate $r$. Consequently, the MISO downlink channel under consideration becomes an ON-OFF channel. In other words, the system could be modelled as a Markov chain with two states, whereby the state $S_0$ represents the ON state, while the state $S_1$ represents the OFF state (see Fig. 2). 

Having modelled the MISO downlink as a Markov chain, we need to find the state transition probabilities as well as the MGFs of service processes during the two states $S_0$ and $S_1$. The matrix $\mathbf{P}$ turns out to be a rank-1 matrix again; and thus, its two entries are given below:
\begin{align}
p_0=e^{\kappa (1-2^{\frac{r}{B}})}, \ \
p_1=1-e^{\kappa (1-2^{\frac{r}{B}})}. \nonumber
\end{align}
The MGFs remain the same as before. Finally, the EC is given as:
\begin{align}
\label{eq:rMISO}
EC= - \frac{1}{\alpha} \ln (e^{\kappa (1-2^{\frac{r}{B}})} e^{- \alpha rT}+1-e^{\kappa (1-2^{\frac{r}{B}})})
\end{align}

{\bf Computation of the optimal transmission rate.}
Once again, for the scenario of no CSI at the BS, the EC of the IRS-aided MISO system could be maximized further by searching for the optimal rate $r^*$ at the BS. This is done by taking the derivative of Eq. \ref{eq:rMISO} w.r.t. $r$ and equating it to zero. Doing so, and after simplification we get the following transcendental equation:
\begin{align}
\frac{r}{B}\ln(2)+\ln(e^{\alpha r T}-1)=\frac{B \alpha T}{ \kappa \ln(2)}.
\end{align}
One potential way to solve above equation is to assume $e^{\alpha r T}-1\approx e^{\alpha r T} $ which in turn implies that $e^{\alpha r T}>>1$. Under this assumption, the optimal transmission rate is given as:
\begin{align}
\label{eq:rastmiso}
    r^\ast\approx\frac{B\alpha T}{\kappa \ln(2)(\frac{\ln(2)}{B}+\alpha T)}.
\end{align}
\section{Simulation Results}
\label{sec:results}

Simulations were performed in Matlab. The values of some of the most significant simulation parameters are provided in Table I (unless otherwise noted). 

\begin{table}[h!]
\centering

 \begin{tabular}{||c c c||} 
 \hline
 Simulation Parameter & Notation & Value \\ [0.5ex] 
 \hline\hline
 Distance between the BS and the IRS & $d_1$ & 50 m  \\ 
 \hline
 Distance between the IRS and the UE & $d_2$ & 50 m  \\
 \hline
 Length of the IRS & $x_{IRS}$ & 1 m \\
 \hline
 Width of the IRS & $y_{IRS}$ & 1 m  \\
 \hline
 Angle of incidence at the IRS & $\varphi$ & $30^{\circ}$  \\
 \hline
 Gain of the antenna at the BS & $G_t$ & 10 dB  \\
 \hline
 Gain of the antenna at the UE & $G_r$ & 10 dB  \\
 \hline
 Transmit power of the BS & $p_t$ & 1 mW  \\
 \hline
 Noise power at the UE & $\sigma^2$ & 1 $\mu$ W  \\
 \hline
 Total number of Transmit antennas (MISO scenario) & $N_t$ & 10  \\[1ex] 
 \hline
\end{tabular}
\caption{Important simulation parameters}
\end{table}

\subsection{SISO scenario}

\begin{figure}[htb!]
\begin{center}
	\includegraphics[width=3.8in]{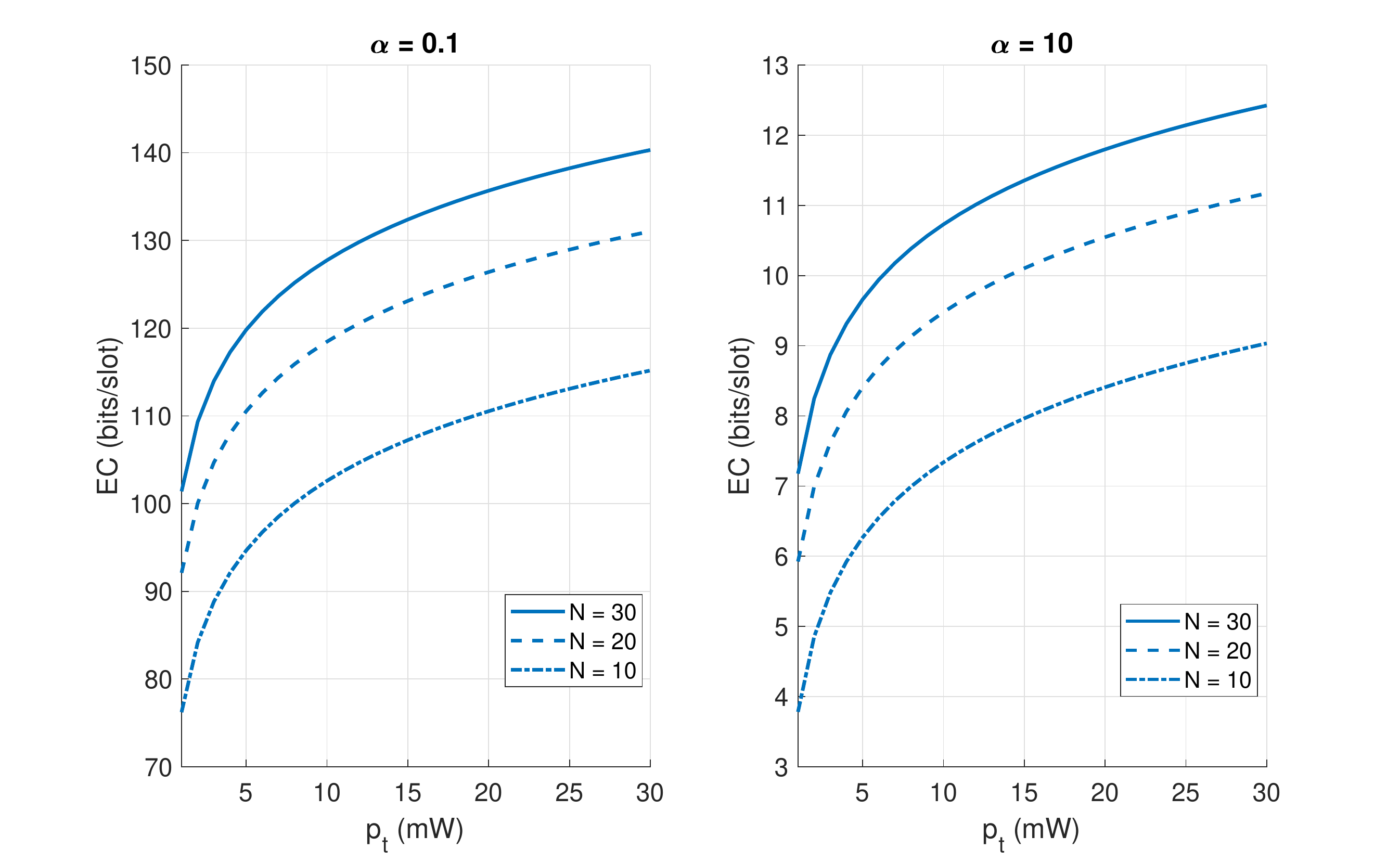} 
\caption{The impact of the number of IRS elements ($N$) on the EC of the SISO downlink for the case of perfect CSI at the BS: delay-tolerant communication regime (left), delay-limited communication regime (right). }
\label{fig:ecvsptkcsi}
\end{center}
\end{figure}

Fig. \ref{fig:ecvsptkcsi} plots the EC of the SISO downlink against the total power budget $p_t$ of the BS when it has the perfect CSI at its disposal. Fig. \ref{fig:ecvsptkcsi} reveals the following: i) there is a logarithmic increase in the EC with the increase in $p_t$ (because ultimately, the EC is a variant of the Shannon's capacity, and is defined as the log of the MGF of the service process); ii) there is also a logarithmic increase in the EC with the increase in the number of IRS elements $N$ (but note that increasing $N$ also increases IRS hardware cost as well as the CSI acquisition overhead at the BS and at the IRS); iii) the EC of the delay-tolerant communication (figure on the left with $\alpha=0.1$) is significantly (more than ten-fold) larger than the EC of the delay-limited communication (figure on the right with $\alpha=10$).

\begin{figure}[htb!]
\begin{center}
	\includegraphics[width=3.8in]{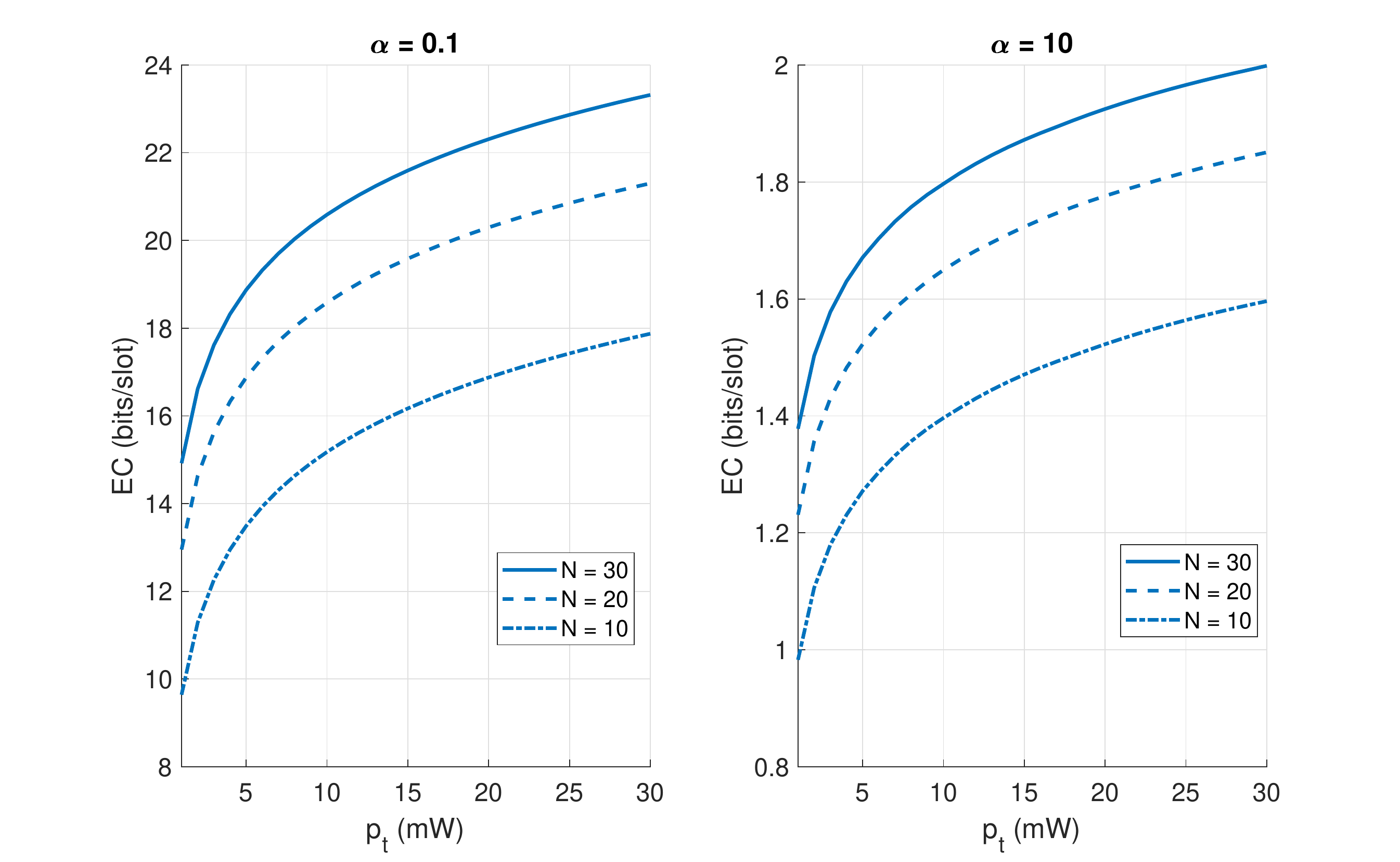} 
\caption{The impact of the number of IRS elements ($N$) on the EC of the SISO downlink for the case of no CSI at the BS: delay-tolerant communication regime (left), delay-limited communication regime (right).}
\label{fig:ecvsptucsi}
\end{center}
\end{figure}

Fig. \ref{fig:ecvsptucsi} is the repeat of Fig. \ref{fig:ecvsptkcsi} for the case when the CSI is not available at the BS. Not only that, the key points learned from the Fig. \ref{fig:ecvsptucsi} are also the same as the Fig. \ref{fig:ecvsptkcsi}. Furthermore, a quick comparison of the two figures (Fig. \ref{fig:ecvsptkcsi} and Fig. \ref{fig:ecvsptucsi}) reveals that the presence of the CSI at the BS has a significant positive impact on the EC (for both delay-tolerant and delay-limited communication regimes). Specifically, the EC for the case of perfectly-known CSI at the BS is at least five times greater than the EC for the case of no CSI at the BS. Note that the optimal transmission rate $r^\ast$ was computed and thereafter utilized to compute the EC for Fig. \ref{fig:ecvsptucsi}.

\subsection{MISO scenario}

\begin{figure}[ht]
\begin{center}
	\includegraphics[width=3.8in]{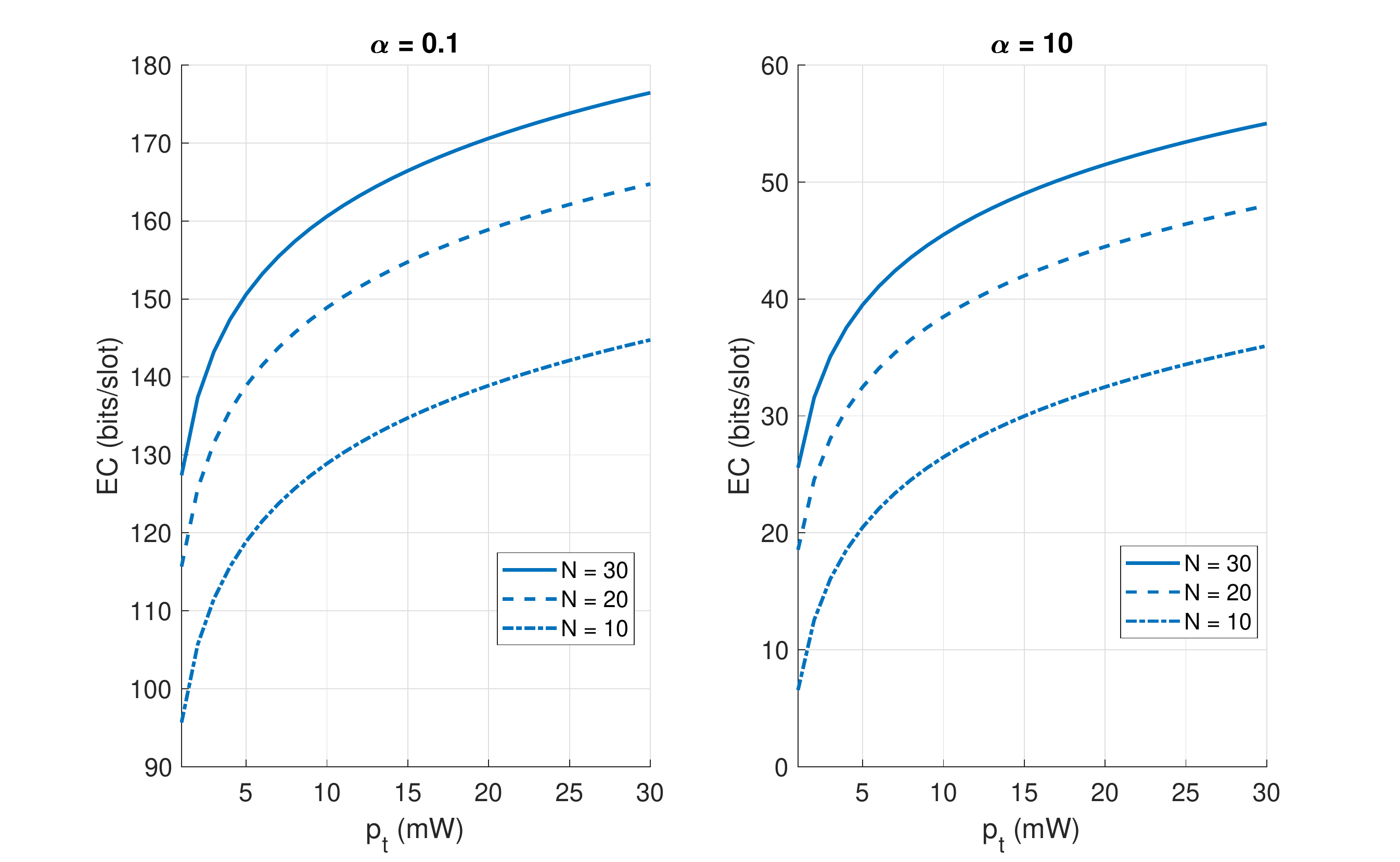} 
\caption{The impact of the number of IRS elements ($N$) on the EC of the MISO downlink for the case of perfect CSI at the BS: delay-tolerant communication regime (left), delay-limited communication regime (right). }
\label{fig:ecvsptMKcsa}
\end{center}
\end{figure}

Fig. \ref{fig:ecvsptMKcsa} plots the EC of the MISO downlink against the total power budget $p_t$ of the BS when it has the perfect CSI at its disposal. Fig. \ref{fig:ecvsptMKcsa} lets us infer similar trends as in Fig. \ref{fig:ecvsptkcsi}. That is, the EC increases logarithmically with the increase in $p_t$ as well as with the increase in the number of IRS elements $N$. Moreover, the EC of the delay-tolerant communication (figure on the left with $\alpha=0.1$) is significantly (at least three times) larger than the EC of the delay-limited communication (figure on the right with $\alpha=10$). Last but not the least, a quick comparison of Fig. \ref{fig:ecvsptMKcsa} with Fig. \ref{fig:ecvsptkcsi} reveals a gain of at least 30 bits/slot in the EC due to the presence of multiple antennas ($N_t=10$, in this case) at the BS, for medium-to-high values of $p_t$. 

\begin{figure}[ht]
\begin{center}
	\includegraphics[width=3.8in]{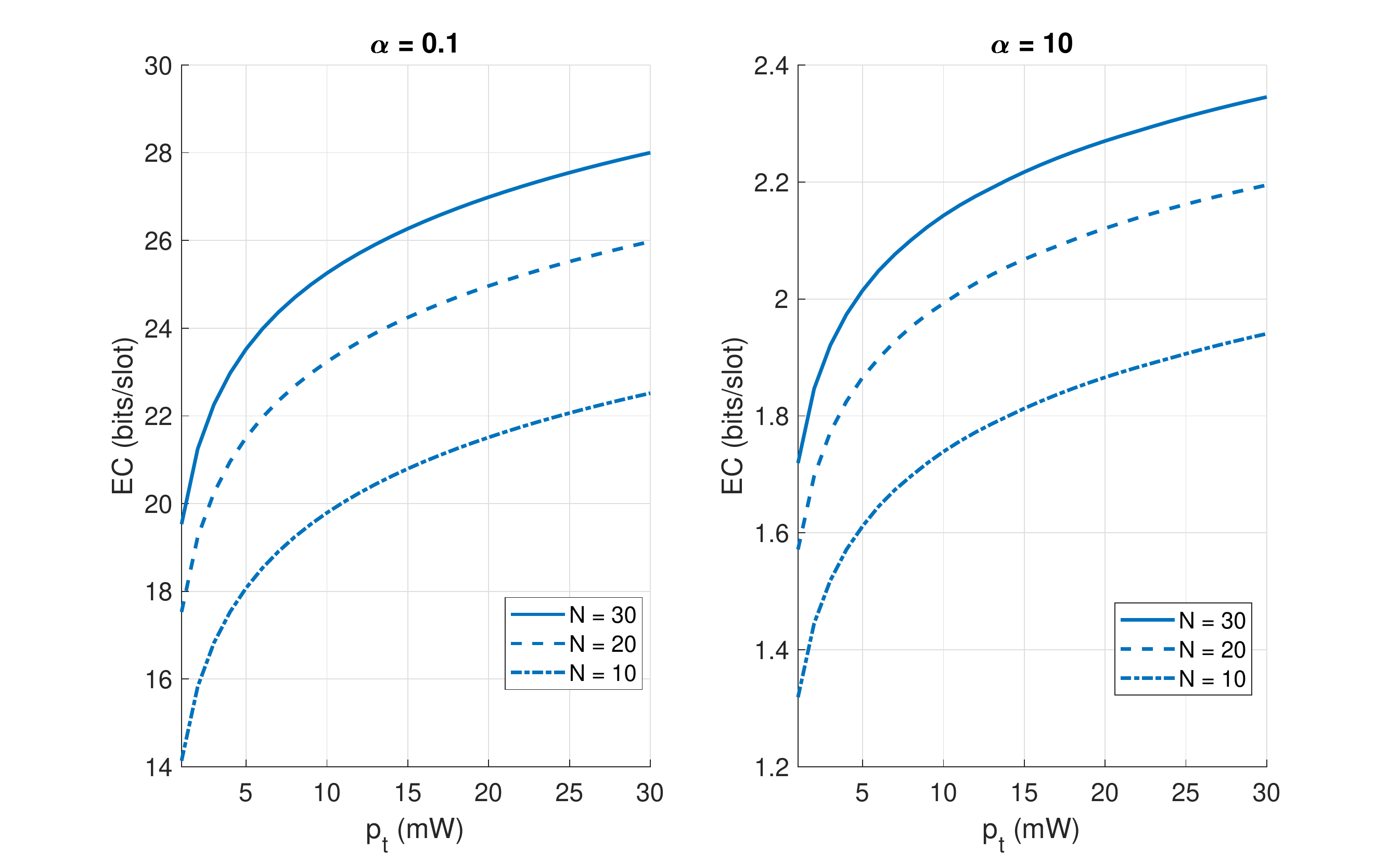} 
\caption{The impact of the number of IRS elements ($N$) on the EC of the MISO downlink for the case of no CSI at the BS: delay-tolerant communication regime (left), delay-limited communication regime (right). }
\label{fig:ecvsptmusci}
\end{center}
\end{figure}

Fig. \ref{fig:ecvsptmusci} is again the repeat of Fig. \ref{fig:ecvsptMKcsa} for the case when the CSI is not available at the BS. Not only that, the key points learned from the Fig. \ref{fig:ecvsptmusci} are also the same as the Fig. \ref{fig:ecvsptMKcsa}. Furthermore, a quick comparison of the two figures (Fig. \ref{fig:ecvsptMKcsa} and Fig. \ref{fig:ecvsptmusci}) reveals that the lack of the CSI at the BS has an adverse effect on the EC (for both delay-tolerant and delay-limited communication regimes). Specifically, the EC reported in Fig. \ref{fig:ecvsptmusci} is at least five times lower than the EC reported in Fig. \ref{fig:ecvsptMKcsa}. Last but not the least, comparing Fig. \ref{fig:ecvsptmusci} with Fig. \ref{fig:ecvsptucsi}, we observe a gain of at least 5 bits/slot in the EC of the MISO downlink compared to the EC of the SISO downlink (for the delay-tolerant communication regime). On the other hand, for the delay-limited communication regime, the gain in the EC of the MISO downlink over the EC of the SISO downlink is negligible, due to very strict QoS constraints. Note once gain that the optimal transmission rate $r^\ast$ was computed and thereafter utilized to compute the EC for Fig. \ref{fig:ecvsptmusci}. 

\begin{figure}[ht]
\begin{center}
	\includegraphics[width=3.8in]{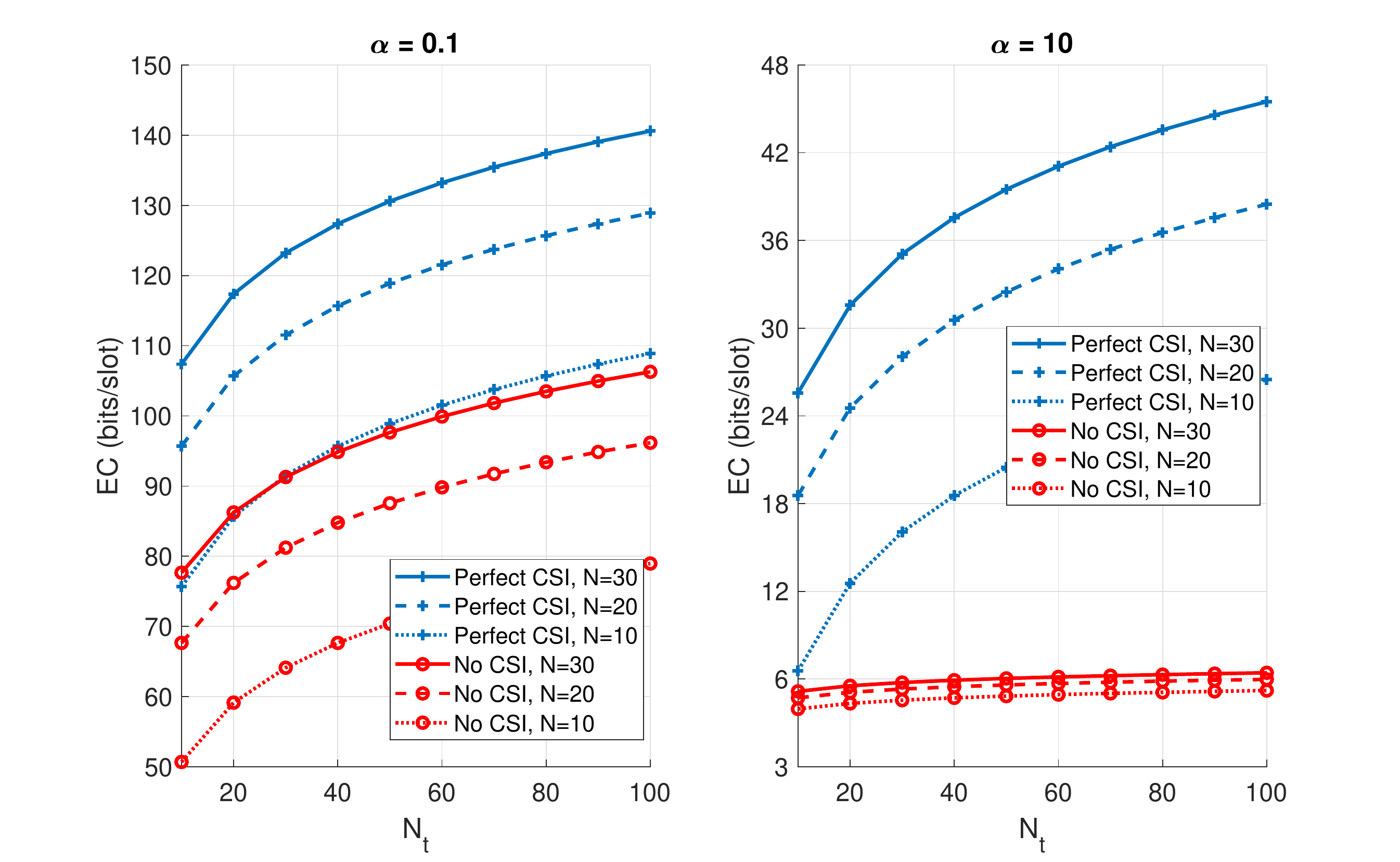} 
\caption{The impact of the number of transmit antennas ($N_t$) at the BS on the EC of the MISO downlink: delay-tolerant communication regime (left), delay-limited communication regime (right). }
\label{fig:ecvsnt}
\end{center}
\end{figure}

Fig. \ref{fig:ecvsnt} plots the EC against the number of transmit antennas $N_t$ at the BS, for the delay-tolerant communication regime with $\alpha=0.1$ (sub-plot on the left) and the delay-limited communication regime with $\alpha=10$ (sub-plot on the right). Moreover, each of the two sub-plots considers the two extreme cases of the availability of the CSI at the BS (i.e., perfect CSI and no CSI) for various values of the number of IRS elements $N$. We first discuss the delay-tolerant communication regime (sub-plot on the left). Here, we observe that the EC for the case of perfect CSI at the BS is at least 30 bits/slot more than the EC for the case of no CSI at the BS, for a given $N$ and and for any value of $N_t$. Moreover, we observe a logarithmic increase in the EC with the increase in an increase in either $N_t$ or $N$. Next, the delay-limited communication regime (sub-plot on the right). Here, the most important finding is that we don't observe any gains in the EC with increase in $N_t$ for the pessimistic case of delay-limited communication with no CSI, as expected. This is because the lack of CSI and very strict QoS requirements ($\alpha=10$) together act like a dual-edge sword which eventually diminish any potential gains in the EC due to increase in either $N_t$ or $N$. 

\subsection{Optimization of transmission rate when CSI is not available at the BS}

\begin{figure}[htb!]
\begin{center}
	\includegraphics[width=3.8in]{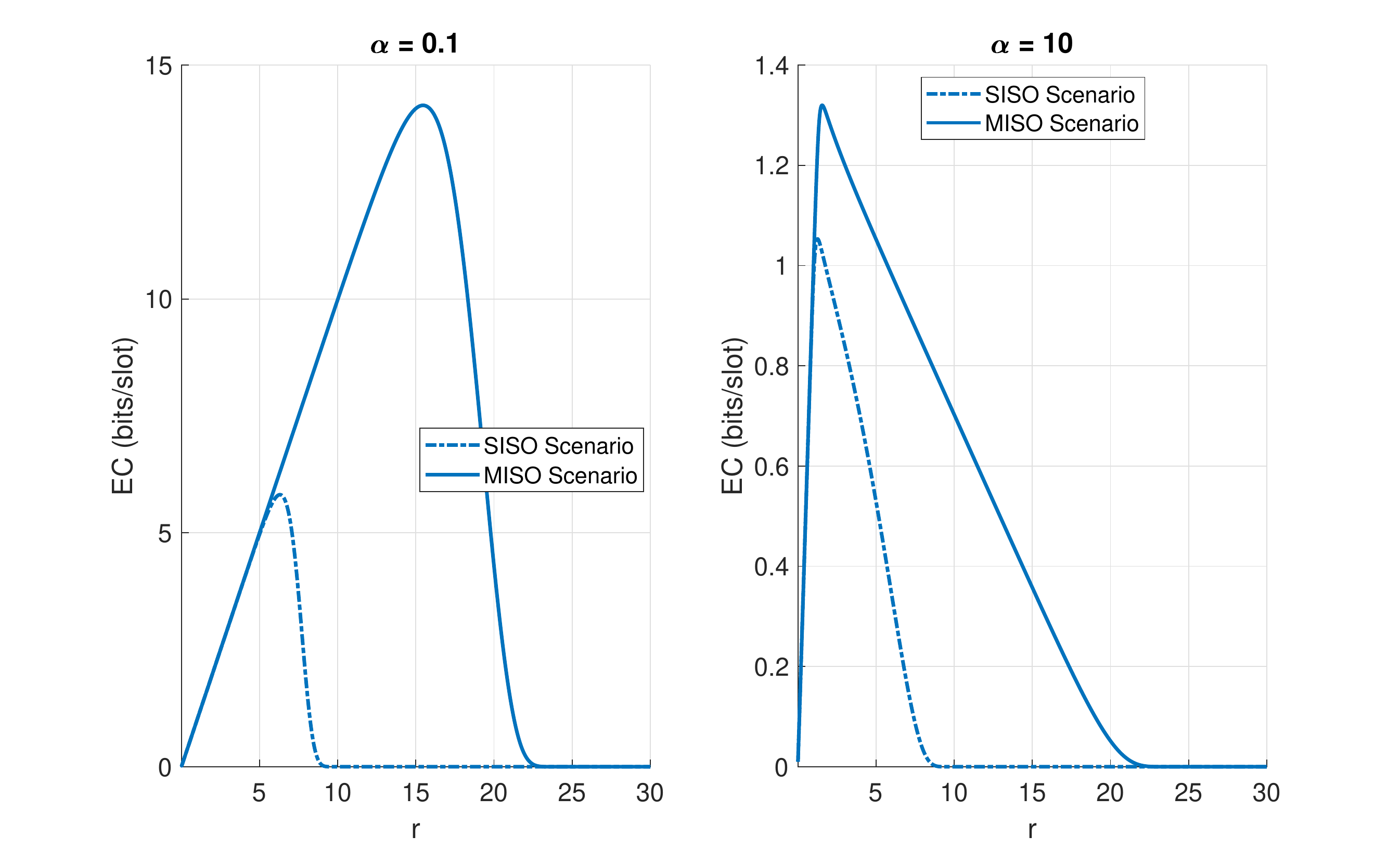} 
\caption{The EC turns out to be a concave-like function of the transmission rate $r$. (The sub-plot on the left depicts the delay-tolerant communication regime, while the sub-plot on the right depicts the delay-limited communication regime.)}
\label{fig:ecvsr}
\end{center}
\end{figure}

Finally, Fig. \ref{fig:ecvsr} plots the EC against the transmission rate $r$ for the delay-tolerant communication regime with $\alpha=0.1$ (see the sub-plot on the left) and the delay-limited communication regime with $\alpha=10$ (see the sub-plot on the right), for the case of no CSI at the BS. Moreover, each of the two sub-plots considers both SISO and MISO scenarios. Fig. \ref{fig:ecvsr} corroborates our conjecture that the EC is a concave-like function of $r$. This revelation allows us to find the optimal transmission rate $r^\ast$ to enhance the EC further for the case of no CSI at the BS. Note that we kept $p_t=1$ mW, and $N_t=10$ (for MISO scenario) to generate Fig. \ref{fig:ecvsr}.

\section{Conclusion}
\label{sec:conclusion}

This work studied the QoS performance of an IRS-assisted SISO/MISO downlink in terms of a quantitative measure---the so-called effective capacity. Closed-form expressions for the EC (for both SISO/MISO downlink) were obtained under two extreme assumptions regarding the availability of the CSI, i.e., perfect CSI at the BS, and no CSI at the BS. Extensive simulations were done, which revealed a number of key points. That is, the EC increases logarithmically with the increase in either of the number of IRS elements, or number of transmit antennas at the BS, or the transmit power budget of the BS. This implies that $N$, $N_t$, $p_t$ help improve the EC to a certain extent, beyond which the EC could only be increased by increasing, say, the bandwidth of the system. Some other alternate potential mechanisms to boost the QoS performance of the system are the following: incorporation of a re-transmission protocol, e.g., automatic repeat request (ARQ) protocol, transmit diversity schemes, relaying-based schemes. Simulation results further revealed that the lack of CSI at the BS and very strict QoS requirements together prove to be a destructive combination which makes any potential gains in the EC due to an increase in either $N_t$ or $N$ void.

This work opens up many exciting opportunities for the future work, some of which are as follows. One could study the EC-based QoS performance of an IRS-assisted SISO/MISO downlink: i) to quantify the increase in the EC of the system when the BS and the UE implement ARQ protocol (and its variants); ii) to reassess the EC of the system for the more realistic scenario of discrete phase shifts by the individual IRS elements; iii) when only imperfect CSI is available at the BS and at the IRS.

\section*{Acknowledgement}
This publication was made possible by the NPRP awards [NPRP12S-0225-190152] and [NPRP10-1231-160071] from the Qatar National Research Fund, a member of the Qatar Foundation. Furthermore, the first author (Waqas Aman) would like to thank the Higher Education Commission (HEC), Pakistan for providing him the opportunity for a research visit to the University of Glasgow, Glasgow, UK under IRSIP scholarship program. The statements made herein are solely the responsibility of the authors. 
\appendices
\section{Proof of Proposition 2.1}

Let $h_i=a_ie^{j\theta_i}$, $g_i=b_ie^{j\psi_i}$ be the polar form representation of $h_i$ and $g_i$, respectively. Note that $a_i$ and $b_i$ are i.i.d. Rayleigh random variables with scale parameter $\nu=1$. Thus, one could rewrite Eq. \ref{eq:y} as follows:
\begin{align}
y=\sqrt{p_t\zeta}\sum_{i=1}^Na_ib_ie^{j(\phi_i+\theta_i+\psi_i)}x + w
\end{align}
Recall from assumption A3 that the IRS controller has perfect CSI of both hops at its disposal, it employs transmit beamforming mechanism which leads to zero phase error at the UE. In other words, the IRS controller sets the phase of its $i$-th element as follows: $\phi_i=-\theta_i-\psi_i$. Thus, the SNR at the UE becomes:
\begin{align}
\gamma=\frac{p_t\zeta\vert \sum_{i=1}^Na_ib_i\vert^2}{\sigma^2}=\left(\frac{\sqrt{p_t\zeta} \sum_{i=1}^N(a_ib_i)}{\sigma}\right)^2
\end{align}

Let $A_i=\frac{\sqrt{p_t\zeta}}{\sigma}a_ib_i$, then $E(A_i)=\frac{\sqrt{p_t\zeta}\pi}{2\sigma }$ and $V(A_i)=\frac{p_t \zeta(16-\pi^2)}{4\sigma^2}$. For $N$ reasonablly large, one could invoke the Central Limit Theorem to get the following result: $\sum_{i=1}^NA_i\sim \mathcal{N}(N(\frac{\sqrt{p_t \zeta}\pi}{2\sigma }),N(\frac{p_t \zeta(16-\pi^2)}{4\sigma^2}))$. Finally,
\begin{align}
\gamma=N\frac{p_t \zeta(16-\pi^2)}{4\sigma^2}\chi_1^2(\lambda).
\end{align}

\section{Proof of Proposition 2.2}
To derive the distribution of the SNR for the IRS-assisted MISO downlink, we begin by assuming $N=2$ and $N_t=2$. This allows us to write the following equation:
\begin{align}
\label{eq:app21}
\mathbf{g}^H\mathbf{\Pi}\mathbf{Hf}=\pi_1g_1(f_1h_{11}+f_2h_{12})+\pi_2g_2(f_1h_{21}+f_2h_{22})
\end{align}
where $\pi_i=e^{j\phi_i}$. We design the reflection coefficients matrix $\mathbf{\Pi}$ at the IRS as follows: $\mathbf{\Pi} = diag[\mathbf{\frac{g}{||g||^2}}]$. Thus, $\mathbf{g}^H\mathbf{\Pi} = [1, \, 1]$. In plain words, we have designed $\mathbf{\Pi}$ such that the IRS does transmit beamforming on the second hop (which effectively cancels out the effects of the channels on the second hop). Therefore, we could re-write Eq. \ref{eq:app21} as follows:
\begin{align}
\label{eq:app22}
\mathbf{g}^H\mathbf{\Pi}\mathbf{Hf}=(f_1h_{11}+f_2h_{12})+(f_1h_{21}+f_2h_{22})
\end{align}

Let $\mathcal{A}=(f_1h_{11}+f_2h_{12})$, and $\mathcal{B}=(f_1h_{21}+f_2h_{22})$. Then, one could verify that $\mathcal{A} \sim C\mathcal{N}(0,\vert f_1\vert^2+\vert f_2\vert^2)$, and that $\mathcal{B} \sim C\mathcal{N}(0,\vert f_1\vert^2+\vert f_2\vert^2)$. Thus, $\mathbf{g}^H\mathbf{\Pi}\mathbf{Hf}$ is the sum of two i.i.d. complex Gaussian random variables (for $N=N_t=2$). In general:
\begin{align}
\mathbf{g}^H\mathbf{\Pi}\mathbf{Hf}=\sum_{i=1}^NZ_i,
\end{align}
 where $Z_i \sim C\mathcal{N}(0,\sum_{j=1}^{N_t}\vert f_j\vert^2)$. Thus, $\mathbf{g}^H\mathbf{\Pi}\mathbf{Hf}$ is the summation of $N$ i.i.d. random variables. For $N$ reasonably large, one could invoke Central Limit theorem to get the following approximation: $\mathbf{g}^H\mathbf{\Pi}\mathbf{Hf}\sim C\mathcal{N}(0,N\sum_{j=1}^{N_t}\vert f_j\vert^2)$. This implies that: $\vert \mathbf{g}^H\mathbf{\Pi}\mathbf{Hf}\vert \sim \text{Rayleigh}  (N\sum_{j=1}^{N_t}\vert f_j\vert^2)$, and $\frac{p_t \zeta\vert \mathbf{g}^H \mathbf{\Pi}\mathbf{Hf}\vert^2}{\sigma^2} \sim \text{exp} (\kappa)$, where $\kappa =\frac{\sigma^4}{2N^2(p_t \zeta\sum_{j=1}^{N_t}\vert f_j\vert^2)2}$.

\footnotesize{
\bibliographystyle{IEEEtran}
\bibliography{main}

\begin{thebibliography}{10}
\providecommand{\url}[1]{#1}
\csname url@rmstyle\endcsname
\providecommand{\newblock}{\relax}
\providecommand{\bibinfo}[2]{#2}
\providecommand\BIBentrySTDinterwordspacing{\spaceskip=0pt\relax}
\providecommand\BIBentryALTinterwordstretchfactor{4}
\providecommand\BIBentryALTinterwordspacing{\spaceskip=\fontdimen2\font plus
\BIBentryALTinterwordstretchfactor\fontdimen3\font minus
  \fontdimen4\font\relax}
\providecommand\BIBforeignlanguage[2]{{%
\expandafter\ifx\csname l@#1\endcsname\relax
\typeout{** WARNING: IEEEtran.bst: No hyphenation pattern has been}%
\typeout{** loaded for the language `#1'. Using the pattern for}%
\typeout{** the default language instead.}%
\else
\language=\csname l@#1\endcsname
\fi
#2}}

\bibitem{ozdogan:WCL:2019}
O.~{Ozdogan}, E.~{Bjornson}, and E.~G. {Larsson}, ``Intelligent reflecting
  surfaces: Physics, propagation, and pathloss modeling,'' \emph{IEEE Wireless
  Communications Letters}, pp. 581--585, May 2019.

\bibitem{Gong:Arxiv:2019}
S.~Gong, X.~Lu, D.~T. Hoang, D.~Niyato, L.~Shu, D.~I. Kim, and Y.-C. Liang,
  ``Toward smart wireless communications via intelligent reflecting surfaces: A
  contemporary survey,'' \emph{IEEE Communications Surveys \& Tutorials},
  vol.~22, no.~4, pp. 2283--2314, 2020.

\bibitem{zhao:arXiv:2019}
J.~Zhao, ``A survey of intelligent reflecting surfaces (irss): Towards 6g
  wireless communication networks,'' \emph{arXiv e-prints}, Nov. 2019.

\bibitem{huang:arXiv:2020}
C.~{Huang}, S.~{Hu}, G.~C. {Alexandropoulos}, A.~{Zappone}, C.~{Yuen},
  R.~{Zhang}, M.~D. {Renzo}, and M.~{Debbah}, ``Holographic mimo surfaces for
  6g wireless networks: Opportunities, challenges, and trends,'' \emph{IEEE
  Wireless Communications}, vol.~27, no.~5, pp. 118--125, Oct. 2020.

\bibitem{wu:CM:2020}
Q.~{Wu} and R.~{Zhang}, ``Towards smart and reconfigurable environment:
  Intelligent reflecting surface aided wireless network,'' \emph{IEEE
  Communications Magazine}, vol.~58, no.~1, pp. 106--112, Jan. 2020.

\bibitem{Huang:TWC:2019}
C.~{Huang}, A.~{Zappone}, G.~C. {Alexandropoulos}, M.~{Debbah}, and C.~{Yuen},
  ``Reconfigurable intelligent surfaces for energy efficiency in wireless
  communication,'' \emph{IEEE Transactions on Wireless Communications},
  vol.~18, no.~8, pp. 4157--4170, Aug. 2019.

\bibitem{Schober:arXive:2019}
X.~{Yu}, D.~{Xu}, and R.~{Schober}, ``{Enabling Secure Wireless Communications
  via Intelligent Reflecting Surfaces},'' \emph{arXiv e-prints}, p.
  arXiv:1904.09573, Apr. 2019.

\bibitem{Wu:GLOBECOM:2018}
Q.~{Wu} and R.~{Zhang}, ``Intelligent reflecting surface enhanced wireless
  network: Joint active and passive beamforming design,'' in \emph{IEEE Global
  Communications Conference (GLOBECOM)}, Dec. 2018, pp. 1--6.

\bibitem{zhao2:ARXiv:2020}
M.-M. Zhao, Q.~Wu, M.-J. Zhao, and R.~Zhang, ``Intelligent reflecting surface
  enhanced wireless network: Two-timescale beamforming optimization,''
  \emph{arXiv e-prints}, Sep. 2020.

\bibitem{Zhou:TSP:2020}
G.~Zhou, C.~Pan, H.~Ren, K.~Wang, and A.~Nallanathan, ``A framework of robust
  transmission design for irs-aided miso communications with imperfect cascaded
  channels,'' \emph{IEEE Transactions on Signal Processing}, vol.~68, pp.
  5092--5106, Aug. 2020.

\bibitem{zhao:ARXiv:2020}
M.-M. Zhao, Q.~Wu, M.-J. Zhao, and R.~Zhang, ``Exploiting amplitude control in
  intelligent reflecting surface aided wireless communication with imperfect
  csi,'' \emph{arXiv e-prints}, May 2020.

\bibitem{Huang:JSAC:2020}
C.~{Huang}, R.~{Mo}, and C.~{Yuen}, ``Reconfigurable intelligent surface
  assisted multiuser miso systems exploiting deep reinforcement learning,''
  \emph{IEEE Journal on Selected Areas in Communications}, vol.~38, no.~8, pp.
  1839--1850, jun. 2020.

\bibitem{Wu:TWC:2003}
D.~Wu and R.~Negi, ``Effective capacity: a wireless link model for support of
  quality of service,'' \emph{IEEE Trans. Wireless Commun.}, vol.~2, no.~4, pp.
  630--643, Jul. 2003.

\bibitem{Gursoy:TWC:2010}
S.~Akin and M.~C. Gursoy, ``Effective capacity analysis of cognitive radio
  channels for quality of service provisioning,'' \emph{IEEE Trans. Wireless
  Commun.}, vol.~9, no.~11, pp. 3354--3364, Nov. 2010.

\bibitem{Anwar:TVT:2016}
A.~H. Anwar, K.~G. Seddik, T.~ElBatt, and A.~H. Zahran, ``Effective capacity of
  delay-constrained cognitive radio links exploiting primary feedback,''
  \emph{IEEE Trans. Veh. Tech.}, vol.~65, no.~9, pp. 7334--7348, Sep. 2016.

\bibitem{gross2012scheduling}
J.~Gross, ``Scheduling with outdated csi: Effective service capacities of
  optimistic vs. pessimistic policies,'' in \emph{Proc. Workshop on Quality of
  Service}, 2012, pp. 1--9.

\bibitem{Gursoy:TIT:2013}
D.~Qiao, M.~C. Gursoy, and S.~Velipasalar, ``Effective capacity of two-hop
  wireless communication systems,'' \emph{IEEE Trans. Info. Theory}, vol.~59,
  no.~2, pp. 873--885, Feb. 2013.

\bibitem{Lateef:TC:2009}
S.~Ren and K.~B. Letaief, ``Maximizing the effective capacity for wireless
  cooperative relay networks with qos guarantees,'' \emph{IEEE Trans. Commun.},
  vol.~57, no.~7, pp. 2148--2159, Jul. 2009.

\bibitem{Soret:TWC:2010}
B.~Soret, M.~C. Aguayo-Torres, and J.~T. Entrambasaguas, ``Capacity with
  explicit delay guarantees for generic sources over correlated rayleigh
  channel,'' \emph{IEEE Trans. Wireless Commun.}, vol.~9, no.~6, pp.
  1901--1911, Jun. 2010.

\bibitem{WShah:WCL:2019}
S.~W.~H. {Shah}, M.~M.~U. {Rahman}, A.~N. {Mian}, A.~{Imran}, S.~{Mumtaz}, and
  O.~A. {Dobre}, ``On the impact of mode selection on effective capacity of
  device-to-device communication,'' \emph{IEEE Wireless Commun. Lett.}, vol.~8,
  no.~3, pp. 945--948, Jun. 2019.

\bibitem{waqas:ICC:2020}
W.~{Aman}, Z.~{Haider}, S.~W.~H. {Shah}, M.~M. {Ur Rahman}, and O.~A. {Dobre},
  ``On the effective capacity of an underwater acoustic channel under
  impersonation attack,'' in \emph{IEEE International Conference on
  Communications (ICC)}, Jul. 2020.

\bibitem{Tang:TWC:2007}
J.~Tang and X.~Zhang, ``Quality-of-service driven power and rate adaptation
  over wireless links,'' \emph{IEEE Transactions on Wireless Communications},
  vol.~6, no.~8, pp. 3058--3068, Dec. 2007.

\bibitem{Murihead:Wiley:1982}
R.~Muirhead, \emph{Aspects of multivariate statistical theory}, ser. Wiley
  Series in Probability and Mathematical Statistics.\hskip 1em plus 0.5em minus
  0.4em\relax John Wiley \& Sons New York, 1982.

\bibitem{Chang:TNCS:2012}
C.~Chang, \emph{Performance Guarantees in Communication Networks}, ser.
  Telecommun. Netw. and Computer Syst.\hskip 1em plus 0.5em minus 0.4em\relax
  Springer London, 2012.

\bibitem{Yu:2012}
Y.~A. Brychkov, ``On some properties of the marcum q function,'' \emph{Integral
  Transforms and Special Functions}, vol.~23, no.~3, pp. 177--182, 2012.

\end{thebibliography}
}

\vfill\break

\end{document}